\documentclass[epj,final]{svjour}
\usepackage{epsfig}

\newcommand{\bmi}[2]{\begin{minipage}[#1]{#2 \linewidth}}
\newcommand{\emi}{\end{minipage}}
\newcommand{\beq}{\begin{equation}}
\newcommand{\eeq}{\end{equation}}
\newcommand{\bea}{\begin{eqnarray}}
\newcommand{\eea}{\end{eqnarray}}
\newcommand{\beas}{\begin{eqnarray*}}
\newcommand{\eeas}{\end{eqnarray*}}
\newcommand{\bdm}{\begin{displaymath}}
\newcommand{\edm}{\end{displaymath}}
\newcommand{\fett}{{\bf\Pi}}
\newcommand{\loopa}{\ensuremath{T H^{2d-1} \int\!\!\int \frac{d^d q d^d r}
{(2\pi)^{2d}}}}

\newcommand{\tkd}{T H^{1+2\epsilon} K_d^2}

\newcommand{\qint}{\int\frac{d^d q}{(2\pi)^d}}

\newcommand{\bpi}{{\ensuremath{\bf\Pi}}}
\newcommand{\cs}{{\ensuremath{\cal S}}}
\newcommand{\cz}{{\ensuremath{\cal Z}}}

\renewcommand{\thesubsection}{\alph{subsection})}

\begin{document}

\title{Low temperature thermodynamics of inverse square spin models
in one dimension}

\author{W. Hofstetter\inst{1}\and W. Zwerger\inst{2}}
\institute{Theoretische Physik III, Universit\"at Augsburg, 86135 Augsburg, Germany\and 
Sektion Physik, Ludwig-Maximilians-Universit\"at\ M\"unchen, Theresienstr.\ 37,
80333 M\"unchen, Germany}

\date{April 1998}

\abstract{
We present a field-theoretic 
renormalization group calculation in two loop order
for classical $O(N)$-models with an inverse square interaction in the vicinity 
of their lower critical dimensionality one. 
The magnetic susceptibility at low temperatures is shown to diverge like 
$T^{-a} \exp(b/T)$ with $a=(N-2)/(N-1)$ and $b=2\pi^2/(N-1)$. 
From a comparison with the exactly solvable Haldane-Shastry model we find
that the same temperature dependence applies also to ferromagnetic quantum 
spin chains. }


\maketitle

\section{Introduction}
The investigation of long range forces in statistical mechanics has 
a considerable history, including studies on the basic requirements
for a proper thermodynamic limit \cite{Griffiths 71} or on the 
existence of phase transitions
even in one-dimensional systems. For classical spins, it has been
known for some time \cite{Ruelle 68} that ferromagnetic models 
with a pair-interaction
decaying like $n^{-(1+\sigma)}$ ($\sigma >0$ is necessary for a proper
thermodynamic limit) exhibit no transition if $\sigma >1$. Choosing 
$\sigma <1$ however, it turns out there is indeed a phase transition
even in one dimension both for Ising models \cite{Dyson 69} as well as for systems
with a continuous symmetry \cite{Froehlich 78}. 
The borderline case $\sigma = 1$ of an inverse
square interaction is thus of particular interest. For Ising spins an early
conjecture by Thouless \cite{Thouless 69} that this model 
exhibits a peculiar continuous
transition with a finite jump of the order parameter was finally proven 
rigorously by Fr\"ohlich and Spencer \cite{Froehlich xx}. 
In the case of a continuous 
symmetry like the $XY$- or Heisenberg-models with inverse square interaction,
Simon has shown \cite{Simon 81} that no symmetry breaking 
appears here, suggesting 
that the borderline interaction exhibiting a Kosterlitz-Thouless type 
transition is $(\ln n)/n^2$. An alternative proof that the inverse square 
$XY$-model in one dimension exhibits no long range order was later 
given by Simanek \cite{Simanek 87}, using the 
classical version of the Mermin-Wagner theorem.

A quantitative calculation of the low temperature susceptibility of long range
ferromagnetic spin models was given by Kosterlitz \cite{Kosterlitz 76} 
within a momentum shell renormalization group (RNG) up to 
one loop order. In particular the susceptibility was found to diverge 
exponentially like $\exp(2\pi^2/(N-1)T)$ in the case $\sigma=1$ 
of an inverse square interaction for models with a continuous 
symmetry, where only the trivial fixed point at $T=0$ 
describing a fully ordered ground state exists. Independently, in a brief
note, Br\'ezin, Zinn-Justin and Le Guillou \cite{Brezin 76} presented the 
results of a two loop field-theoretic RNG of the $O(N)$-model with interaction
proportional to $n^{-(1+\sigma)}$. They determined the critical exponents
up to order $(d-\sigma)^2$ in the vicinity of the lower critical dimension
$d_c=\sigma$. Here we present a detailed version of the field-theoretic 
two loop calculation which essentially follows their ideas. However, it
uses a somewhat different method to determine the renormalization 
factors and, moreover, concentrates on the behaviour in the marginal 
case $d=\sigma=1$.

The study of one-dimensional classical spin models with long range interactions
is interesting not only from a pure statistical mechanics point of view,
but is also relevant for actual physical problems. In fact the inverse 
square Ising-model is closely related to the well-known Kondo-problem
of a magnetic impurity locally coupled to the conduction electron 
spin density \cite{Anderson 70}. 
A more recent application of the same model is the two
state system with ohmic dissipation, arising e.g. in the problem of 
quantum coherence between macroscopically different states 
\cite{Leggett 87,Chakravarty 95}. 
For the inverse square $XY$-case it turns out that the problem of strong
tunneling in a metallic single-electron-box can be treated by calculating
the free energy of the classical spin model in the presence of twisted 
boundary conditions \cite{Hofstetter 97}. 
It was the latter problem which motivated our present work, whose intention
is in fact twofold: First of all, we give a detailed account of the 
two-loop RNG of the $O(N)$-model with inverse square interaction which 
has obviously not been done in the literature so far. In addition to that
we discuss the corresponding $S=1/2$ quantum spin chain. In particular, 
it is found that the low temperature behaviour of the susceptibility 
is essentially identical with that for classical spins.

\section{RNG in two-loop order}
\subsection{Model and Definitions}
We start with a model of classical unit spins ${\bf S}(n)$ with $N$ components 
which are placed on a ring with $L$ sites, thus enforcing periodic 
boundary conditions. The interaction energy in an external magnetic field
$H$ along the $N$-direction is given by 
\beq
H_{cl} = -\sum_{n\ne n'} [d(n-n')]^{-2}\,{\bf S}(n)\,{\bf S}(n') - 
H\sum_n S^N (n)
\eeq
with 
\beq
d(n) = \frac{L}{\pi} \sin\frac{\pi |n|}{L}
\eeq
the chord distance between spins which are $n$ sites apart. The associated 
partition function can then be written as 
\beq  \label{generating} \textstyle
\mathcal{Z} = \int \prod_{n}d{\bf S}(n)\; \delta\!\left({\bf S}^2(n) -1\right) 
\exp\lbrace-S(T,H,{\bf S}(n))\rbrace.   
\eeq 
with an action 
\beq    \label{euklidwirkung}
\mathcal{S} =\frac{1}{2T} \sum_{n\ne n'} \frac{\left({\bf S}(n) - {\bf S}(n')
\right)^2}{(d(n-n'))^2} - \frac{H}{T} \sum_n S^N(n).
\eeq
Here $\int d{\bf S}(n) \delta\!\left({\bf S}^2(n) -1\right)$ denotes an integration
over all orientations of a classical $N$-component unit vector ${\bf S}(n)$. 
As is known from the work cited above, this model exhibits a continuous
phase transition to an ordered phase below a critical temperature 
$T_c \sim \epsilon /(N-1)$ for noninteger dimensionality $d=1+\epsilon$ ($\epsilon > 0$).
In order to study the behavior near the lower critical dimensionality
one, we can therefore parametrize the spin ${\bf S}$ according to 
a low temperature expansion 
\beq \label{lowtemp}
{\bf S}(x)=\left(\fett(x), \sqrt{1-\fett^2(x)}\right).
\eeq
The $\Pi_i$ $(i=1,\ldots,N-1)$ are Goldstone modes corresponding to 
the fluctuations around the ordered state with $S_N = 1$ where all 
spins point in the $N$-direction. 
The generating functional is now given by
\beq
\mathcal{Z} = \int\prod_n \frac{d\fett(n)}{\sqrt{1-\fett^2(n)}}\,\exp\lbrace-S(T,H,\fett(n))\rbrace .  
\eeq
The term in the integration measure 
\beq
\prod_n \frac{1}{\sqrt{1-\fett^2(n)}} = \exp\left\{-\frac{1}{2}
\sum_n \ln\left(1-\fett^2(n)\right)\right\}   
\eeq
is due to the $\delta$-function constraint and ensures the $O(N)$-symmetry
of the model, which appears to be broken by the above parametrization.
In a continuum notation $\sum_n \to a^{-d}\int d^d x$ with $a$ the lattice 
constant and 
\beq
\prod_n \frac{1}{\sqrt{1-\fett^2(n)}} = 
\exp\left\{-\frac{1}{2}\:
a^{-d}\int d^d x \ln\left( 1-{\bf\Pi}^2(x)\right)\right\}
\eeq
In the following, we will work in the dimensional regularization
scheme, where 
\beq 
a^{-d}=\int^{\Lambda}\frac{d^d q}{(2\pi)^d} = 0.
\eeq
We can therefore neglect the measure term (see \cite{Amit 84}).

In order to generate correlation functions of the $\Pi$ fields, 
we introduce additional source fields according to 
\beq
\cs \mapsto \cs -\int dx \:{\bf J}(x)\: {\bf \Pi}(x).
\eeq
The quantities which are most conveniently calculated diagrammatically
are the irreducible vertex functions $\Gamma^{(n)}$.  Their generating functional 
$\Gamma[\bar{\bpi}, H]$ is formally obtained from $\ln \cz[{\bf J}, H]$ 
by a Legendre transformation with respect to ${\bf J}$ where $\bar{\bpi}$
is the corresponding classical field. In our calculation, we will use
the two point function
\beq
\Gamma^{(2)}(x,y) = \frac{\delta^2 \Gamma}{\delta \bar{\Pi}(x) 
\delta \bar{\Pi}(y)}.
\eeq
As in the case of the non-linear $\sigma$-model with next-neighbour 
coupling \cite{Amit 84,Zinn-Justin 96} it can be shown that the model 
is renormalizable by introducing two renormalization factors
$Z_t$ and $Z_\pi$ rescaling the temperature and the field ${\bf \Pi}$. 
These factors must be adjusted in such a way that the renormalized 
$N$-point functions 
\beq   \label{renorm_gamma}
\Gamma^{(N)}_R = Z_\pi^\frac{N}{2} \: \Gamma^{(N)}
\eeq
remain finite in terms of the renormalized parameters 
\beq \label{renorm_temp}
t = \kappa^\epsilon Z_t^{-1} T
\eeq
\beq \label{renorm_h}
h = Z_t^{-1} Z_\pi^{\frac{1}{2}} H.
\eeq
Below we will see that the choice $Z_t=Z_\pi\equiv Z$ with a
single independent renormalization constant $Z$ is possible. 

\subsection{Details of the two-loop-calculation}\label{details}

\begin{figure}[t]
\begin{center}
\epsfig{file=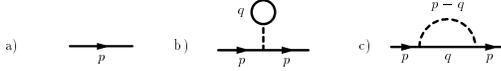,width=0.15\linewidth,angle=90,bbllx=500,bblly=274,
bburx=579,bbury=784}
\end{center}
\caption{\label{fig:one_loop}Free propagator (a)) and one-loop 
contributions to $\Gamma^{(2)}$ (b) and c)).}
\end{figure}

\begin{figure}[t]
\begin{center}
\epsfig{file=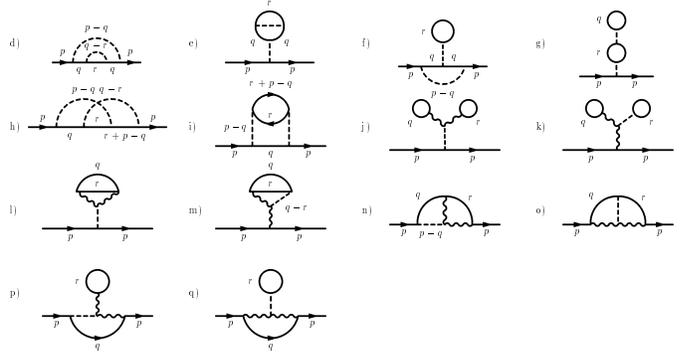,width=0.55\linewidth,angle=90,bbllx=187,bblly=59,
bburx=582,bbury=790}
\end{center}
\caption{\label{fig:two_loop}Two-loop contributions to $\Gamma^{(2)}$.}
\end{figure}

In order to calculate the $Z$-factors introduced above 
to the order $t^2$, we first rewrite the action
in momentum representation:
\bea   \label{reskalwirk}
\cal{S} & = & \frac{1}{T} \qint\, (|q|+H)\, \Big[ \frac{\fett(q)\fett(-q)}{2}
 +\frac{\fett^2(q) \fett^2(-q) }{8}  \nonumber \\ 
&&+ \frac{\fett^2(q) \left(\fett^2\right)^2(-q) }{16} 
\Big] + O\left(\left(\fett^2\right)^4\right), 
\eea
where we have rescaled $H \mapsto 2\pi H$ and $T \mapsto (2\pi)^d T$. 
\begin{figure}[t]
\begin{center}
\epsfig{file=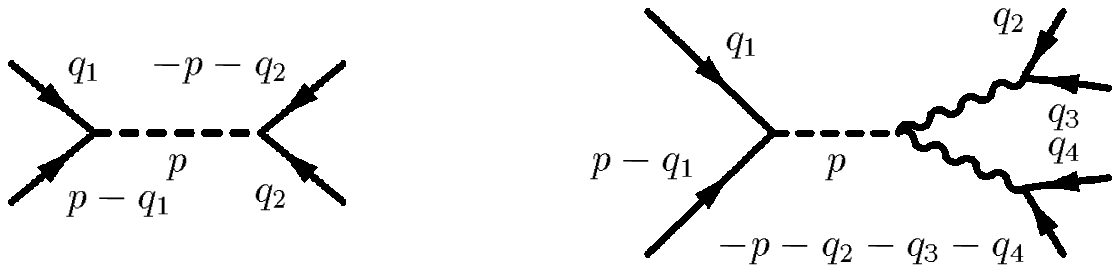, width=0.2\linewidth, angle=90,bbllx=480,bblly=331,
bburx= 580,bbury=791}
\end{center}
\caption{\label{fig:interactions} Diagrammatic representation of the
four- and six-point interaction terms.}
\end{figure}
The free propagator corresponding to the quadratic term has the
form 
\bea
\langle \Pi^\alpha(p) \Pi^\beta(q) \rangle & = & \delta_{\alpha\beta} \,
\delta(p+q) \,  G_0^\alpha(p) \\  
& = &   (2\pi)^d \delta_{\alpha\beta}
\, \delta(p+q) \, \frac{T}{|p| + H}.    \nonumber
\eea
with $\alpha, \beta=1,\ldots, N-1$.
In addition, $\mathcal{S}$ contains two interaction terms which
contribute up to two-loop order. These are shown in 
fig.\ (\ref{fig:interactions})
and correspond to a four- and a six-point-interaction. We will renormalize
the model using the two-point function $\Gamma^{(2)}$. The relevant 
diagrams up to two-loop order are shown in fig.\ (\ref{fig:one_loop})
and (\ref{fig:two_loop}).
In a short hand notation, the two point function in terms of these
diagrams is given by
\bea
\Gamma^{(2)}(k)  
& = &  a) + b) + c) - d) -e) -f) -g) -h) -i)  \nonumber \\ 
& &  +j) +k) +l) +m) +n) +o) +p) +q). 
\eea
Each one of these diagrams is proportional to $T^{L-1}$ where
$L$ is the number of loops. Several of them contain divergences
for $\epsilon\to0$.
In the minimal subtraction scheme, we want to absorb
these divergences in the two renormalization constants
$Z_t$ and $Z_\pi$. After renormalization, $\Gamma_R^{(2)}$ should 
be an analytic function of the variables $h$, $t$ and $p$.

For the action (\ref{reskalwirk}) we note that
the $p$-dependent parts of the diagrams are finite. 
To one-loop order, for instance, only the second diagram (labeled by c)) depends on the 
external momentum
\begin{equation}\label{oneloopterm}
\int \frac{d^d q}{(2\pi)^d} \;\frac{1+|p-q|}{1+|q|} = 
\int \frac{d^d q}{(2\pi)^d} \;\frac{|p-q|-|q|}{1+|q|} .
\end{equation}
Evidently this integral is finite for $d<2$. Because of
\beq
\Gamma_R^{(2)}(p) = \frac{Z_\pi}{Z_t}\:
\kappa^\epsilon\:\frac{|p| + Z_t\,Z_{\pi}^{-\frac{1}{2}}\, h }{t} 
+ \textrm{higher order terms}, 
\eeq  
the coefficient of $|p|$ in the inverse propagator is renormalized by
$Z_\pi/Z_t$. As is evident from the one-loop term (\ref{oneloopterm}),
this prefactor is already finite. Therefore the choice
\beq  \label{zpigleichzt}
\displaystyle
Z_\pi = Z_t \equiv Z    
\eeq
is possible, leaving only one independent renormalization constant.
In fact this relation is valid to arbitrary loop order, as was shown by Br\'ezin,
Zinn-Justin and Le Guillou \cite{Brezin 76}.
As a consequence, we can take the external momentum $p$ to be zero
in the following. 
One further simplification is due to the fact that 
in dimensional regularization ``empty'' integrals of the type
$\int d^d q$ are equal to zero. This implies that the contributions 
of the diagrams in c), h), n) and p) vanish.
Moreover, several contributions turn out to be identical and therefore cancel:
this applies to the diagrams d) and o) which are given by 
\beq 
\loopa \;\frac{1+|q-r|}{(1+|q|)(1+|r|)}.  
\eeq 
Similarly the diagrams i) and m) which give a contribution 
\beq
\frac{N-1}{2} \loopa \;\frac{1+|q-r|}{(1+|q|)(1+|r|)} 
\eeq
are equal and therefore cancel in $\Gamma^{(2)}$.
The remaining expression to be evaluated is 
\beq \label{restgamma}
\Gamma^{(2)}(0, H) = a) + b) - e) - f) - g) + k) + j) + l) + q).
\eeq
In order to extract the singular parts of the integrals, 
we use the asymptotic expansions
\beq\label{identity 1}
\int \frac{d^d q}{(2\pi)^d} \;\frac{1}{1+|q|} 
= K_d \left(-\frac{1}{\epsilon} + o(\epsilon)\right) 
\eeq
and
\beq\label{identity 2}
\int \frac{d^d q}{(2\pi)^d} \;\frac{1}{(1+|q|)^2} = 
K_d \left(1+o\left(\epsilon^2\right)\right) 
\eeq
where the constant 
\beq
K_d = \frac{2}{(4\pi)^{d/2} \Gamma(d/2)}
\eeq
results from the $d$-dimensional angular integration.
The identities (\ref{identity 1}) and (\ref{identity 2}) allow to extract
the leading singularities in $\epsilon$ for all contributions 
to (\ref{restgamma}) except e). 
By a detailed asymptotic analysis of the corresponding integral, 
the following singular contributions of this diagram can be obtained:
\bea
e) &=& \int\!\int \frac{d^d q d^d r}{(2\pi)^{2d}} \frac{1+|r-q|} {(1+|q|)^2 (1+|r|)} \\
&=& \frac{N-1}{2} \tkd \left(\frac{1}{2\epsilon^2} + \frac{1}{2\epsilon}
+ o(1) \right).  \nonumber
\eea 
Neglecting nonsingular parts, the resulting two-point function is
therefore finally given by
\bea
\Gamma^{(2)}(0, H) &=& \frac{H}{T} - H^{1+\epsilon} \frac{N-1}{2\epsilon}   
+ T H^{1+2\epsilon} \frac{(N-1)(N-2)}{4\epsilon} \nonumber \\
&&+ T H^{1+2\epsilon} \frac{3 (N-1)^2}{8\epsilon^2} 
\eea
where we have absorbed the angular integration term $K_d$ in the 
temperature according to
\beq   \label{reskalII}
T \mapsto T K_d^{-1} \ \ ,\ \  \Gamma^{(2)} \mapsto \Gamma^{(2)} K_d.
\eeq
At this stage we can now introduce renormalized parameters 
$h$ and $t$ as defined in  (\ref{renorm_temp}) and (\ref{renorm_h}).
Taking $\kappa=1$ for convenience, we obtain
\bea
\Gamma_R^{(2)}(0,h) &=& \frac{Z^{\frac{1}{2}}\,h}{t} - \frac{N-1}{2\epsilon}
Z^{1+\frac{1+\epsilon}{2}}\,h^{1+\epsilon} \\
&&+ t\,h^{1+2\epsilon} \frac{(N-1)(N-2)}{4\epsilon}  
+ t\,h^{1+2\epsilon}\,\frac{3 (N-1)^2}{8\epsilon^2} \nonumber 
\eea
where terms of order $t^2$ have been neglected. Using the ansatz
\beq  \label{ansatz}
Z = 1 + a_1(\epsilon) t + a_2(\epsilon) t^2 + o(t^3)
\eeq
this can be written as 
\bea  \label{langer_ausdruck}
\lefteqn{\Gamma_R^{(2)}(0,h) = \frac{h}{t} + h \left(\frac{1}{2}a_1 - 
\frac{N-1}{2 \epsilon}\right)+ t\,h \bigg(\frac{4\,a_2-a_1^2}{8}}   \nonumber \\
&&- \frac{(N-1)((3+\epsilon)a_1-(N-2))}{4\epsilon} 
+ \frac{3(N-1)^2}{8\epsilon^2}\bigg) \nonumber \\
&&+ t\,\ln h \bigg(  \frac{3(N-1)^2}{4\epsilon} - 
\frac{(N-1)}{2} \frac{(3+\epsilon)}{2} a_1 \bigg) 
\eea 
where we have neglected terms which are not singular in $\epsilon$.
The coefficients $a_1$ and $a_2$ are now chosen such as to  
eliminate the poles of order $\epsilon^{-1}$ and $\epsilon^{-2}$. 
This is achieved by 
\beq
a_1 = \frac{N-1}{\epsilon}
\eeq
and
\bea
a_2 &=& \frac{(N-1)^2}{\epsilon^2} + \frac{(N-1)}{2\epsilon}, 
\eea
thus rendering the renormalized two point function finite.
The renormalization constant $Z$ is therefore given by
\beq \label{Zresult}
Z = 1 + \frac{N-1}{\epsilon}\, t + \left(\frac{(N-1)^2}{\epsilon^2} +
\frac{N-1}{2\epsilon}\right) t^2 + O(t^3). 
\eeq
From this result we can now determine the beta function
\beq
\beta(t) \equiv - \kappa \frac{\partial t}{\partial \kappa}
\eeq
measuring the variation of the renormalized coupling under
variation of the momentum scale, keeping the bare parameters $T$ and $H$
fixed.
According to (\ref{renorm_temp}), we can write
\beq  \label{betaausz}
\beta(t) = -\epsilon t\left(1 + t \frac{d\ln Z}{dt}\right)^{-1}.
\eeq
Using the result (\ref{Zresult}) this immediately leads to
\beq  \label{finalbeta}
\beta(t) = - \epsilon t + (N-1) t^2 + (N-1) t^3 + O(t^4). 
\eeq
For $\epsilon >0$, the renormalization flow has a non trivial 
fixed point corresponding to a phase transition at finite 
temperature. The associated critical behaviour has been
discussed by Br\'ezin et al.\ \cite{Brezin 76}. Here we are interested
in the one-dimensional case, where only the trivial fixed 
point at $t=0$ describing the fully ordered ground state exists.
During the calculation we have rescaled the coupling $T$ twice.
Inserting the original temperature according to $T \to T/2\pi^2$
we get the following equation for the flow of $T$ under the reduction
of the cutoff $\Lambda\to\Lambda e^{-l}$:
\beq
\frac{dT}{dl} = (N-1)\,\frac{T^2}{2\pi^2}  + 
(N-1)\,\frac{T^3}{(2\pi^2)^2} + O(T^4).
\eeq
Integrating this equation between an initial temperature $T$ very close to zero 
and a final value $T=O(1)$, the corresponding length rescaling factor 
behaves like 
\beq \label{xiresult}
\xi(T) \sim T^{\frac{1}{N-1}}\,\exp\left(\frac{2\pi^2}{(N-1)T}\right).
\eeq
Obviously $\xi(T)$ plays the role of a ``correlation length'', however it is
important to point out that there can be no exponential decay in models
with long range interactions. In fact, using Lieb-Simon-type inequalities, 
it can be shown that the spin-spin correlation function 
\beq
g(n) = \langle {\bf S}(n)\,{\bf S}(0) \rangle
\eeq
asymptotically decays like $n^{-2}$ for arbitrary temperatures \cite{Zwerger 91,Spohn 98}.
In spite of the absence of an exponential decay, within a standard scaling hypothesis 
the behavior of $g(n)$ near the critical point at $T=0$ is described by a universal 
function $f(n)$ with $f(0)=1$ and $f(|n|\to\infty) = f_\infty/n^2$, $f_\infty = O(1)$, 
such that 
\beq
g(n,T\to 0) = f\left(\frac{n}{\xi(T)}\right)
\eeq
with $\xi(T)$ given by (\ref{xiresult}). As a result the magnetic susceptibility
\beq
\chi = \beta \left(1+2\,\sum_{n=1}^\infty g(n) \right)
\eeq
is finite for any $T$ and scales like
\beq \label{suszrenorm}
\chi(T) \sim T^{-\frac{N-2}{N-1}}\, \exp\left(\frac{2\pi^2}{(N-1)T}\right).
\eeq
at low temperatures. The two loop calculation thus uniquely determines both the 
exponent and the $T$-dependence of the prefactor of the magnetic susceptibility
as $T\to 0$.

\section{Discussion}
In this work we have calculated the low temperature susceptibility of 
classical $O(N)$ spin models with an inverse square interaction.
Now it is interesting to compare our results 
with the corresponding ferromagnetic \emph{quantum} spin models.
In the case of a Heisenberg model 
\beq \label{spinhamiltonian}
\hat{H} = - \sum_{n\ne n'} J(n-n')\,\hat{\mathbf{S}}_n \hat{\mathbf{S}}_{n'}
- H \sum_n \hat{S}_n^z
\eeq 
with interaction $J(n) = d^{-2}(n)$, the exact low temperature susceptibility
in the case $S=1/2$ was calculated by Haldane \cite{Haldane 91}. It is given by
\beq \label{exact_susc}
\chi(T) = \frac{1}{4\sqrt{\pi}}\, T^{-\frac{1}{2}} \exp\left(\frac{\pi^2}{4T}\right).
\eeq
and thus exhibits precisely the same temperature dependence as our result 
(\ref{suszrenorm}) for the classical case with $N=3$ (note that we have 
chosen units in which $J=1$ and our temperature differs from the one in 
ref.\ \cite{Haldane 91} by a trivial factor two arising from 
$\sum_{n\ne n'} = 2\sum_{n<n'}$). In fact the only difference is in the numerical 
factor which appears in the exponent. On a qualitative level the exponential 
divergence of $\chi(T)$ in the case of long range interactions can be 
understood from a simple spin wave calculation around the perfectly 
ordered ferromagnetic ground state. For a general spin $S$, the Hamiltonian
(\ref{spinhamiltonian}) leads to a dispersion relation of magnons which 
has the form 
\beq \label{magnondispersion}
\epsilon(k) = 4S\sum_{n>0} J(n)\,(1-\cos(kn)) + H.
\eeq
An equilibrium distribution of noninteracting magnons leads to the standard
reduction in the dimensionless magnetization
\beq
m(T,h) = 1 - \frac{1}{S} \int\frac{dk}{2\pi} \frac{1}{e^{\beta \epsilon(k)} - 1}.
\eeq
For the Haldane-Shastry model with $J(n) = d^{-2}(n)$, the magnon dispersion 
(\ref{magnondispersion}) is given by
\beq
\epsilon(k) = S (2\pi|k| - k^2) + H
\eeq
(see also \cite{Haldane 91}).
The associated reduction in magnetization in the limit $H,T\to 0$ then 
behaves like 
\beq
\delta m(T,H) = \frac{T}{2(\pi S)^2} \ln \frac{2 \pi S}{H} .
\eeq
The condition $\delta m = b$ with a constant $b$ of order one 
then defines a field $H(T)$ 
which qualitatively determines the zero field susceptibility via
\beq
\chi(T) \approx \frac{1}{H(T)} \sim \exp \frac{2 b (\pi S)^2}{T} .
\eeq
Choosing $b=1/2$, 
this result agrees precisely with the exact solution (\ref{exact_susc})
for $S=1/2$, while the exponent in the susceptibility of the classical
spin model with unit spin is obtained by formally setting $S=1$. 
A modified spin wave theory and Schwinger boson techniques  
have been applied to the Haldane-Shastry model by 
Nakano and Takahashi \cite{Nakano 94a,Nakano 94b}.

For short range interactions where $J(n)$ decays faster than $n^{-3}$, 
the magnon dispersion has the usual form 
\beq
\epsilon(k\to0) = Dk^2 + H.
\eeq
It is then straightforward to see that the above spin wave calculation 
leads to a susceptibility diverging like $\chi(T)\sim S^2/T^2$. This 
behavior is indeed found from the exact solution of both the nearest
neighbor $S=1/2$ case \cite{Takahashi 89} as well as for the corresponding
classical Heisenberg model \cite{Fisher 64}. We thus expect that for any
ferromagnetic spin model with a fully ordered ground state the temperature
dependence of the susceptibility at low $T$ is independent of whether
the spins are classical or quantum mechanical.

In the case of an $XY$-model ($N=2$), the ground state for the $S=1/2$ 
chain with inverse square interaction has again a very simple structure, 
as shown by Haldane \cite{Haldane 88}. It would thus be interesting 
to check whether in this case the low temperature susceptibility 
again diverges like that of the classical model with $N=2$, 
i.e. $\chi(T)\sim \exp(2\pi^2/T)$.


\begin{thebibliography}{99}
\bibitem{Simanek 87} E. Simanek, Phys. Lett. {\bf 119A}, 477 (1987).
         Note, however, that the prediction in this paper of an infinite
         susceptibility at low temperatures is not correct 
         as is shown here.
\end{thebibliography}
\end{document}